\documentclass[prl,twocolumn,superscriptaddress,nobibnotes,letterpaper]{revtex4-1}

\usepackage[pdftex]{graphicx}
\usepackage[pdftex]{epsfig}
\usepackage{amsmath}
\usepackage{amssymb}
\usepackage{amsfonts}
\usepackage{color}
\usepackage{wrapfig}
\usepackage{eucal}
\usepackage{hhline}
\usepackage{threeparttable}
\usepackage{supertabular}
\usepackage{multirow}
\usepackage{tabularx}

\usepackage{soul}

\usepackage{array} % make column bigger (vertical)
\usepackage[colorlinks=true, allcolors=blue]{hyperref} %hyperrefs

\newcommand{\und}[1]{_\textrm{#1}}

\definecolor{cgreen}{rgb}{.1,.6,.1}
\definecolor{co}{rgb}{.1,.6,.6}
\definecolor{orange}{rgb}{.9,.4,.0}

\newcolumntype{C}[1]{>{\centering\arraybackslash}p{#1}}

\begin{document}
\pagenumbering{arabic}

\title{Gallium phosphide as a piezoelectric platform for quantum optomechanics}

\author{Robert Stockill}\thanks{These authors contributed equally to this work.}
\affiliation{Kavli Institute of Nanoscience, Department of Quantum Nanoscience, Delft University of Technology, 2628CJ Delft, The Netherlands}
	
\author{Moritz Forsch}\thanks{These authors contributed equally to this work.}
\affiliation{Kavli Institute of Nanoscience, Department of Quantum Nanoscience, Delft University of Technology, 2628CJ Delft, The Netherlands}

\author{Gr\'{e}goire Beaudoin}
\affiliation{Centre de Nanosciences et de Nanotechnologies, CNRS, Universit\'{e} Paris-Sud, Universit\'{e} Paris-Saclay, C2N, 91767 Palaiseau, France}

\author{Konstantinos Pantzas}
\affiliation{Centre de Nanosciences et de Nanotechnologies, CNRS, Universit\'{e} Paris-Sud, Universit\'{e} Paris-Saclay, C2N, 91767 Palaiseau, France}

\author{Isabelle Sagnes}
\affiliation{Centre de Nanosciences et de Nanotechnologies, CNRS, Universit\'{e} Paris-Sud, Universit\'{e} Paris-Saclay, C2N, 91767 Palaiseau, France}

\author{R\'{e}my Braive}
\affiliation{Centre de Nanosciences et de Nanotechnologies, CNRS, Universit\'{e} Paris-Sud, Universit\'{e} Paris-Saclay, C2N, 91767 Palaiseau, France}
\affiliation{Universit\'{e} de Paris, Sorbonne Paris Cit\'{e}, 75207 Paris, France}

\author{Simon Gr\"oblacher}
\email{s.groeblacher@tudelft.nl}
\affiliation{Kavli Institute of Nanoscience, Department of Quantum Nanoscience, Delft University of Technology, 2628CJ Delft, The Netherlands}

%\date{\today}

\begin{abstract}
Recent years have seen extraordinary progress in creating quantum states of mechanical oscillators, leading to great interest in potential applications for such systems in both fundamental as well as applied quantum science. One example is the use of these devices as transducers between otherwise disparate quantum systems. In this regard, a promising approach is to build integrated piezoelectric optomechanical devices, that are then coupled to microwave circuits. Optical absorption, low quality factors and other challenges have up to now prevented operation in the quantum regime, however. Here, we design and characterize such a piezoelectric optomechanical device fabricated from gallium phosphide in which a 2.9~GHz mechanical mode is coupled to a high quality factor optical resonator in the telecom band. The large electronic bandgap and the resulting low optical absorption of this new material, on par with devices fabricated from silicon, allows us to demonstrate quantum behavior of the structure. This not only opens the way for realizing noise-free quantum transduction between microwaves and optics, but in principle also from various color centers with optical transitions in the near visible to the telecom band.
\end{abstract}

\maketitle

The interaction of light and mechanical motion in nanofabricated resonators provides a flexible interface between telecom photons and long-lived phononic modes. Rapid progress in this field has resulted in the realization of nonclassical states of light and motion at the single photon and phonon level~\cite{Riedinger2016,Hong2017,Riedinger2018,Marinkovic2018}, demonstrating the suitability of these structures as quantum light-matter interfaces. One particularly interesting application for which these interfaces could provide their unique functionality is the transduction of quantum information between different carriers. To this end, piezoelectric materials are of great interest, as the electromechanical coupling in principle allows for transduction of a quantum state between the microwave and optical frequency domains~\cite{Bochmann2013,Balram2016,Forsch2018}. Additionally, wide-bandgap materials make simultaneous coupling to both visible wavelength light (at which many optically-active quantum systems operate) and the low-loss telecom bands in the near-infrared possible~\cite{Hill2012,Liu2013}. One of the main challenges to realize such a quantum transducer is the ability to faithfully exchange excitations between optical and mechanical modes, which requires any stray absorption of light to be minimal in order not to introduce thermal noise. 

A particularly interesting class of optomechanical resonators for quantum interfaces are formed from the simultaneous confinement of light and mechanical motion in periodically patterned nanobeams \cite{Eichenfield2009b}. These devices feature high frequency (few GHz) mechanical modes, such that the resolved sideband regime is accessible with reasonable optical resonator quality factors ($\gtrsim2\cdot10^4$), while the mechanical mode can be initialized to the quantum groundstate by cryogenic cooling. The low mass of the mechanical mode and small optical cavity mode volume allow for strong optomechanical coupling and the monolithic design facilitates on-chip integration with other quantum systems. While optomechanical crystals have been fabricated out of materials such as GaAs~\cite{Balram2016,Forsch2018,Ramp2019}, GaP~\cite{Schneider2019,Ghorbel2019}, LiNbO$_3$~\cite{Jiang2019}, SiN~\cite{Davanco2014} AlN~\cite{Bochmann2013,Vainsencher2016} and diamond~\cite{Burek2016}, amongst others, to date nonclassical optomechanical interaction in such structures has been limited to silicon-based devices.

In this work we realize an optomechanical crystal (OMC) in gallium phosphide (GaP) featuring high cooperativity interaction between an optical resonance at 1550~nm and a mechanical breathing mode close to 3~GHz. Due to the minimal absorption of GaP at these wavelengths, cooling our device to $~7$~mK allows us to operate deep in the mechanical quantum groundstate with mode occupations of as little as 0.04 phonons. Furthermore, we demonstrate quantum behavior of our device by measuring nonclassical correlations between photons and phonons~\cite{Riedinger2016}. Our results validate that the device performs at a similar level to comparable designs in Si~\cite{Chan2011,Hong2017,Riedinger2018,Marinkovic2018}, while far surpassing current achievements in GaAs or other piezoelectric materials~\cite{Forsch2018,Ramp2019}. Owing to the wide electronic bandgap and piezoelectric properties of GaP the successful operation of our device in this parameter regime opens the door for novel quantum experiments as well as the potential for using such devices for microwave-to-optics conversion. Strong optical $\chi^{(2)}$ and $\chi^{(3)}$ non-linear coefficients~\cite{Rivoire2009,Wilson2018} and the ability to integrate GaP with current silicon technologies~\cite{Schneider2019} makes this material a unique platform for quantum experiments and technologies.

\begin{figure}
\includegraphics[width = 1\columnwidth]{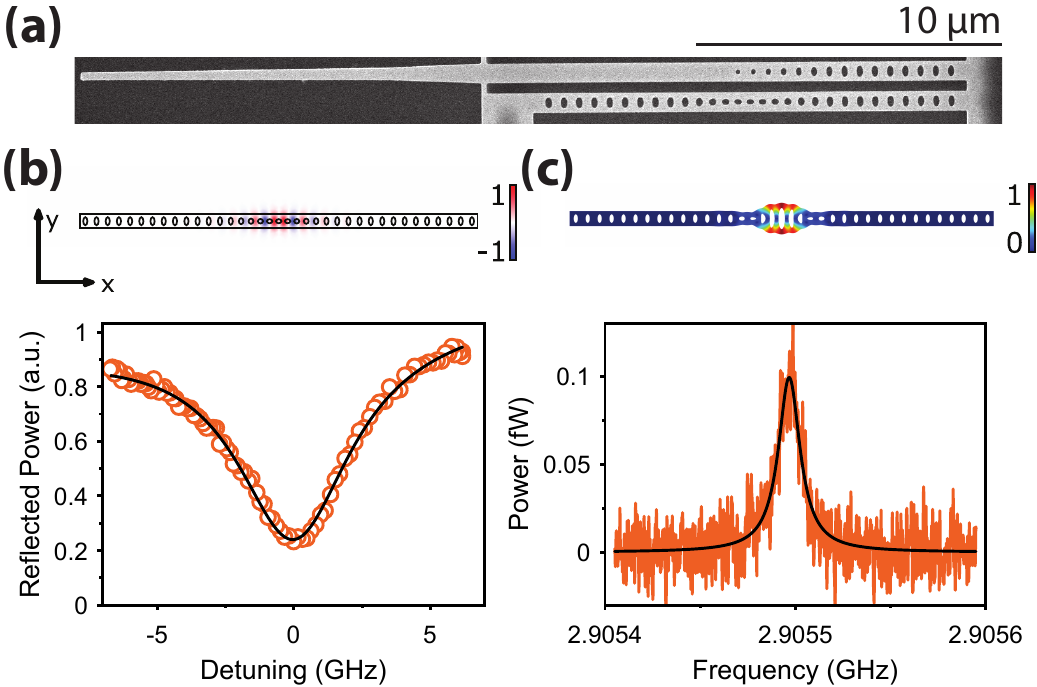}
\caption{\textbf{a} Scanning electron micrograph of the optomechanical crystal (bottom) with a coupling waveguide (top). \textbf{b} Normalized $E\und{y}$ component of the simulated optical mode (top) and corresponding reflection spectrum of the fabricated device (bottom). The solid black curve is a lorentzian fit to the data, including a linear offset to account for changing laser power. The over-coupled linewidth of the optical resonance is $\kappa=2\pi\times5.14$~GHz. \textbf{c} Normalized displacement of the simulated mechanical mode (top) and power spectral density of the mode at $\sim$7mK (bottom). The solid black curve is a lorentzian fit to the data. The linewidth of the resonance is $\gamma\und{m}=2\pi\times$13.8~kHz.}
\label{fig1}
\end{figure}

Our device consists of an optomechanical crystal evanescently coupled to an optical waveguide, as shown in Fig.~\ref{fig1}(a). We fabricate the device from a 200~nm thick layer of GaP, on a 1~$\mu$m Al$_{0.64}$Ga$_{0.36}$P sacrificial layer, both epitaxially grown on a GaP substrate. The material features a large refractive index n$\und{GaP}$ = 3.05 at $\lambda$ = 1550~nm~\cite{Bond1965} as well as a piezoelectric response $\epsilon_{14}=-0.1$~Cm$^{-2}$~\cite{Nelson1968}. The optomechanical device is designed to exhibit an optical mode in the telecom band ($\lambda\sim$1560~nm) as well as a co-localized mechanical mode at $\omega\und{m}\sim2\pi\times2.85$~GHz. The simulated mode profiles are shown in Fig.~\ref{fig1}(b) and (c). The optomechanical coupling between these modes is realized through the photoelastic effect and the moving boundary conditions due to the shape of the mechanical mode \cite{Balram2014}. Simulations of these contributions, with photoelastic coefficients from reference~\cite{Mytsyk2015}, predict a single photon optomechanical coupling strength $g_0=2\pi\times525$~kHz.

\begin{figure*}
\includegraphics[width = 1\textwidth]{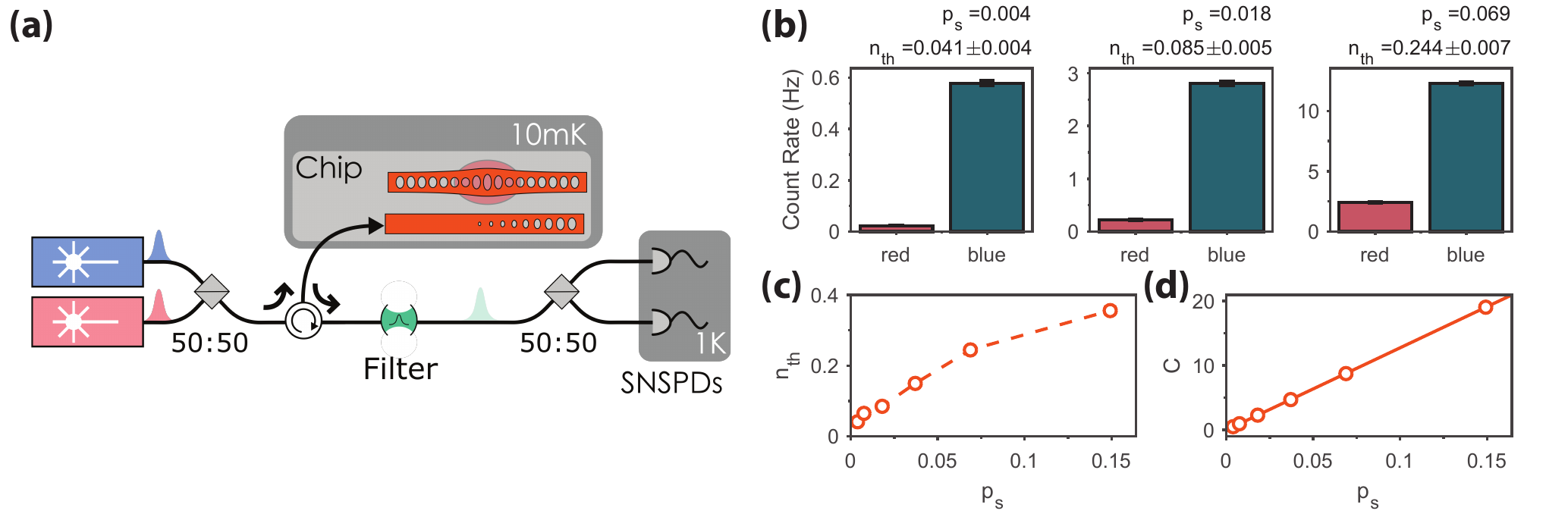}
\caption{Device characterization at Millikelvin temperatures. \textbf{a} Schematic of the measurement setup. We excite or read out our device using a laser on either the red or blue sideband, resulting in cavity-resonant scattered photons. We then filter out the residual pump light and detect the scattered photons on single photon detectors (SNSPDs). \textbf{b} Sideband thermometry measurements taken at varying optical powers (corresponding to different scattering probabilities). The bars represent integrated counts over the duration of the pulse and the error bars correspond to one standard deviation. \textbf{c} Extracted thermal occupations from the sideband thermometry measurement. In this panel, the errorbars are smaller than the corresponding data points. \textbf{d} Extracted values of the cooperativity $C$ as a function of the scattering probability. The line is a linear fit to the data.}
\label{fig2}
\end{figure*}

We place our sample inside a dilution refrigerator at a base temperature of $\sim$7~mK. Optical access to the device is provided by lensed fiber coupling to the evanescantly coupled reflective waveguide (see Fig.~\ref{fig1}(a)). The initial characterization is performed by scanning a tunable laser across the optical resonance at $\omega\und{c} = 2\pi\times194.8$~THz, shown in Fig.~\ref{fig1}(b). The mechanical resonance at $\omega\und{m}=2\pi\times2.905$~GHz is obtained by monitoring GHz-frequency noise in the reflected light, when stabilizing the laser a few GHz blue-detuned from the optical resonance, shown in Fig.~\ref{fig1}(c). We extract optical and mechanical linewidths of  $\kappa=2\pi\times5.14$~GHz (loaded Q-factor of $3.79\times10^{4}$) and $\gamma\und{m}=2\pi\times13.8$~kHz, respectively (cf.\ Fig.~\ref{fig1}(b,c)). The optical resonator is over-coupled to the nearby waveguide (see Supplementary Information for more details), such that we extract an intrinsic linewidth of $2\pi\times1.31$~GHz (intrinsic Q-factor of $1.49\times10^{5}$). 

While the cryogenic cooling of the device should initialize the mechanical mode in its quantum groundstate, previous research~\cite{Forsch2018,Ramp2019} has indicated that, owing to optical absorption, the measured thermal occupation can be significantly higher~\cite{Meenehan2014}. We therefore experimentally determine the absorption-limited mode temperature using sideband thermometry~\cite{Diedrich1989,Safavi-Naeini2012}. In order to limit the effects of absorption-induced heating, we send 40~ns long optical pulses to our device, spaced by 160~$\mu$s. During this measurement, we stabilize our drive laser to either the blue ($\omega\und{l}=\omega\und{c}+\omega\und{m}$) or red ($\omega\und{l}=\omega\und{c}-\omega\und{m}$) sideband of the optomechanical cavity. The blue (red) pulses realize a two-mode squeezing (state-swap) interaction, which creates (annihilates) an excitation of the mechanical mode and produces a scattered photon on resonance with the optical cavity. We then filter out the reflected pump light and use superconducting nanowire single photon detectors (SNSPDs) to measure the photons on cavity resonance. The corresponding setup is shown in Fig.~\ref{fig2}(a). In the weak excitation limit, the photon rates resulting from the blue ($\Gamma\und{b}$) and red ($\Gamma\und{r}$) sideband drives are proportional to $p\und{s}\left(n\und{th}+1\right)$ and $p\und{s}$n$\und{th}$, respectively. Here, $p\und{s}$ is the optical power-dependent scattering probability~\cite{Riedinger2016} (see SI for details). This measurement allows us to extract the thermal occupation of the mechanical mode, $n\und{th}$, using the ratio n$\und{th}=\Gamma\und{r}/(\Gamma\und{b}-\Gamma\und{r})$. By performing this measurement at several optical powers, we can assess the thermal mechanical occupation due to quasi-instantaneous heating from the optical drive, which forms the baseline for further experiments. A selection of photon count rates is shown in Fig.~\ref{fig2}(b) and the full set of extracted thermal occupations is shown in Fig.~\ref{fig2}(c) as a function of $p\und{s}$. The measured thermal occupations are as low as $n\und{th} = 0.041\pm0.004$, a significant improvement over the lowest reported thermal occupations in other III/V-based optomechanical devices~\cite{Forsch2018,Ramp2019}. From the asymmetry measurements, we extract the single-photon optomechanical coupling strength, $g_0=2\pi\times845\pm34$~kHz. We attribute the discrepancy between the simulated and measured values of $g_0$ to our imperfect knowledge of the photoelastic coefficients of GaP, as well as small fabrication-related differences between the designed and actual devices. The power-dependent optomechanical cooperativity, $C \equiv 4g^2/\kappa\gamma\und{m}$, with $g = \sqrt{n\und{c}}g\und{0}$ and $n\und{c}$ being the intra-cavity photon number, is shown in Fig.~\ref{fig2}(d). Operating at the highest cooperativities comes at the cost of a raised thermal occupation. However, cooperativities $C\gg$~1 are achievable while remaining well below $n\und{th}=1$. In particular, $C$ exceeds 1 for a low  occupation of $n\und{th}<0.1$, an important benchmark for the conversion of quantum states~\cite{Hill2012}.

\begin{figure}
\includegraphics[width = 1\columnwidth]{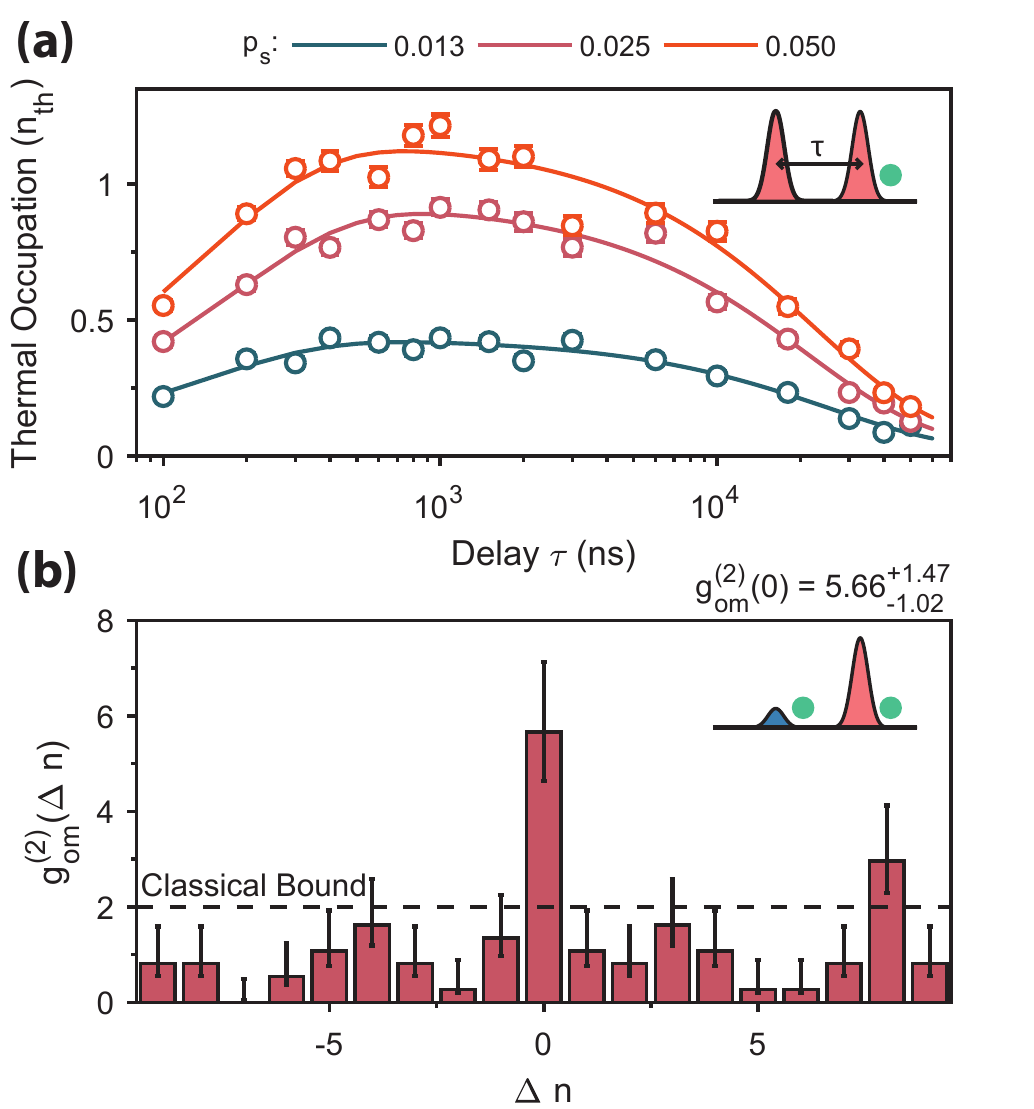}
\caption{\textbf{a} Heating dynamics of a GaP optomechanical crystal. The data points display the measured occupation of the mechanical mode following an initial red-detuned state swap pulse, which results in scattering probabilities of 0.013, 0.025 and 0.050. The occupation is inferred from the count rate of a second red-detuned pulse, delayed by time $\tau$, as displayed in the inset schematic. The solid curves are phenomenological two-exponential fit to the data, accounting for the delayed onset of thermal excitation and the relaxation time of the mechanical mode. \textbf{b} Correlations between state-projecting and state readout photons. The displayed two-photon correlations are calculated by normalizing to the single photon rates during the blue-(red-)detuned write (state-swap) pulse. The errorbars represent 68\% confidence intervals.}
\label{fig3}
\end{figure}

The parameter regime in which our device operates is similar to recent demonstrations of nonclassical behavior in silicon optomechanical resonators~\cite{Riedinger2016,Hong2017,Riedinger2018,Marinkovic2018}. In these optomechanical devices delayed heating of the mechanical mode is a major challenge for quantum experiments~\cite{Forsch2018}. In order to study the presence of similar effects in our current system, we investigate the heating dynamics by sending two consecutive red detuned pulses to the device. The first pulse provides a source of heating to the thermalized device. The second pulse then converts any thermal phonons at an elevated occupation into photons, resulting in an increased scattering rate ($\Gamma\und{r}\propto\textnormal{n}\und{th}$). By varying the delay between the two pulses, we can access the time dynamics of the heating process. The measured mode occupations are shown in Fig.~\ref{fig3}(a) for three different scattering probabilities. To gain insight into the heating dynamics, the occupations are fitted with a phenomenological biexponential function ($n\und{th}\left(\tau\right)=A\cdot e^{-\tau/\tau\und{decay}}\cdot (1-e^{-\tau/\tau\und{rise}})+n\und{th,i})$. We extract the characteristic decay constant, $\tau\und{decay}$ for the mechanical mode to be $22\pm1$~$\mu$s, in reasonable agreement with the Q-factor extracted from the linewidth of the mechanical mode shown in Fig.~\ref{fig1}(c). For these scattering probabilities, we recover a heating time for the mode $\tau\und{rise}$, between 150 and 180 ns.

In order to investigate the suitability of GaP as a piezolectric material platform for quantum optomechanics, we use a pulsing scheme from the DLCZ protocol~\cite{Duan2001,Riedinger2016}. It consists of a weak blue-detuned pulse, which excites the mechanical mode with a small probability and returns a scattered photon on resonance with the cavity. A subsequent red-detuned pulse swaps the state of the mechanical mode onto the light field and again produces a cavity-resonant scattered photon. Just as before, both the excitation and readout pulse are 40~ns long. Based on the characteristic heating time, $\tau\und{rise}$, we choose the time delay between our excitation and readout pulses to be 150~ns. As such, we read out the state of the mechanical mode before the delayed heating of the mode adds excess thermal population.  Given the small probability of exciting the device, detection of a scattered photon from the weak blue-detuned pulse heralds the creation of a nonclassical mechanical excitation of the resonator, consisting predominantly of a single phonon~\cite{Hong2017}. The correlation of the projection photon and the remaining phonon (accessed through the red-detuned state-swap pulse) can then be determined through the statistics of the detected photons during many repetitions of the sequence. We operate at scattering probabilities for the write and read pulses of $p\und{s,write}\sim$0.06\% (corresponding to 25~nW peak power) and $p\und{s,read}\sim$2\% (750~nW), respectively, to limit the parasitic heating, and set the repetition rate to be 25~kHz, such that the mechanical mode can fully re-thermalize between each sequence. More details on the experimental parameters can be found in the S.I..

We analyze the second order correlation $g\und{om}^{(2)}(\Delta\textnormal{n})=P(W\cap R,\Delta\textnormal{n})/(P(W)\times P(R))$ between detector clicks originating from write (W) and read (R) pulses from the same ($\Delta\textnormal{n}=0$) or different ($\Delta\textnormal{n}\neq 0$) pulse sequences. Here, $P(W\cap R,\Delta\textnormal{n})$ is the probability of detecting a read photon $\Delta\textnormal{n}$ sequences after a write photon, while $P(W)$ and $P(R)$ are the independent detection probabilities of write- and read photons. The measured correlation values are displayed in Fig.~\ref{fig3}(b). We observe a correlation of $g\und{om}^{(2)} = 5.66_{-0.98}^{+1.51}$ for read and write pulses from the same sequence, and confirm that detection events from different sequences are on-average uncorrelated. The uncertainty in our measured correlation value is a 68\% confidence interval determined from the likelihood function of rare two-photon coincident events occurring. The high levels of bunching are a clear signature of nonclassical phonon-photon correlations in this system. The correlations we measure here are limited by three sources -- the presence of residual incoherent heating in the device from both the write and read pulses, dark-counts in the SNSPDs (0.08~Hz, $\sim$1/5 of the write pulse clicks), which are important during the low-probability write pulse, and the imperfect filtering of the drive laser in the detection setup. In order to limit the effect of readout-induced heating and detector dark-counts, we consider events occurring in the first 50\% of the read pulse, and the central 50~ns of the write pulse, respectively. Small improvements to our optomechanical device and setup will hence allow to realize more complex single phonon experiments~\cite{Hong2017,Riedinger2018}.

Based on the optomechanical performance shown in this report and the piezoelectric properties of GaP~\cite{Nelson1968}, it is pertinent to consider how the device presented here could be adapted for the conversion of quantum states between the microwave and optical domains. This could be achieved by supplementing the nanobeam with a mechanically coupled, resonant piezoelectric actuator~\cite{Wu2019}. Faithful conversion of a quantum state requires that less than one photon of noise is added throughout such a conversion process. At the same time, efficient conversion requires that the cooperativity of the electromechanical interface must be much greater than 1 and closely matched to the optomechanical cooperativity~\cite{Hill2012}. As a result of the low optical absorption in GaP observed in our measurements, for an optomechanical cooperativity above 10 (see Fig.~\ref{fig2}(c)-(d)), a matched electromechanical interface could result in an added noise figure of only 0.02 photons owing to incoherent heating of the device and imperfect sideband resolution~\cite{Zeuthen2016}.

Maintaining a large optomechanical coupling strength requires for the piezoelectric resonator to have a similar mass to the 280~fg mechanical mode studied here. Considering the piezoelectric tensor of GaP and the alignment of the nanobeam to the [110] crystal orientation, the appropriate breathing mode could then be actuated through a vertical electric field. A generic resonator of matched mass and frequency can be formed by designing a 950~nm wide and 580~nm long beam between superconducting electrodes in a parallel-plate capacitor configuration~\cite{Bochmann2013} (see SI). The resulting interface is expected to convert electrical and mechanical energy with a coupling coefficient $k\und{eff}^2\approx0.017\%$. Assuming a mechanical loss rate equal to our nanobeam, and taking pessimistic values for on-chip parasitic capacitance of $\sim$100~fF, an electromechanical cooperativity of 20 could be achieved by the addition of a microwave-resonator with a Q-factor of as little as $\sim$150. While further investigations of microwave losses in GaP are required, the devices presented here, together with the recently developed GaP-on-Silicon platform~\cite{Wilson2018} offer robust materials systems for incorporation of low-loss coplanar microwave circuitry.

We have demonstrated nonclassical behavior of an optomechanical crystal fabricated from gallium phosphide. Our device can be operated deep in the quantum groundstate with significantly reduced heating compared to similar devices fabricated from GaAs and other piezoelectric materials. GaP combines several unique properties, such as a large electronic bandgap, high refractive index, and a significant piezoelectric response, making it extremely well-suited for many novel applications for optomechanical quantum systems. Combined with the piezoelectric properties of GaP, our demonstration of nonclassical optomechanical interaction will enable microwave-to-optics converters to operate in a previously unreached regime. In addition, the large bandgap could allow for coupling of photons to quantum systems which natively operate in the visible spectrum. Our system can further be used for non-linear optics experiments~\cite{Rivoire2009,Lake2016} owing to the strong optical $\chi^{(2)}$ and $\chi^{(3)}$ non-linearities and the large index contrast between the suspended device and vacuum, potentially even allowing to increase the optomechanical coupling rate into the strong-coupling regime~\cite{Lemonde2016}.

\begin{acknowledgments}
We would like to thank Vikas Anant, Kees van Bezouw, Claus G\"artner, Igor Marinkovi\'{c}, Richard Norte, Amir Safavi-Naeini, and Andreas Wallucks for valuable discussions and support. We also acknowledge assistance from the Kavli Nanolab Delft. This work is supported by the French RENATECH network, as well as by the Foundation for Fundamental Research on Matter (FOM) Projectruimte grants (15PR3210, 16PR1054), the European Research Council (ERC StG Strong-Q, 676842), and by the Netherlands Organization for Scientific Research (NWO/OCW), as part of the Frontiers of Nanoscience program, as well as through a Vidi grant (680-47-541/994).
\end{acknowledgments}

\bibliography{Mirror}

\setcounter{figure}{0}
\renewcommand{\thefigure}{S\arabic{figure}}
\setcounter{equation}{0}
\renewcommand{\theequation}{S\arabic{equation}}

\clearpage

\section{Supplementary Information}

\subsection{Device fabrication}

The samples were fabricated from an epitaxial structure grown by MOCVD on a GaP $\langle 100\rangle$ substrate, consisting of a 1~$\mu$m thick sacrificial layer of Al$_{0.64}$Ga$_{0.36}$P, followed by a 200~nm thick device layer of GaP. The growth was performed in a Veeco Turbodisc D180 reactor under hydrogen as carrier gas, trimethylgallium and trimethylaluminum as organometallic precursors and under phosphine at a reactor pressure of 70 Torr. The OMC pattern is first defined in a $\sim$200~nm thick layer of AR-P 6200-13 resist using a Raith 5000+ direct-write electron beam writer. The pattern is then transferred into the device layer using a reactive ion etch process in an Alcatel GIR-300 etcher using a N$_2$/BCl$_3$/Cl$_2$ chemistry. The remaining resist is removed in a 80$^{\circ}$C bath of N-N Dimethyl Formamide (DMF). The OMC structures are suspended by selectively removing the sacrificial layer using a 10\% Ammonium Fluoride (NH$_4$F) solution. To avoid stiction between the devices and the substrate layer, we use critical point drying as a final step before a 5~nm thin film of AlO$\und{x}$ is deposited using atomic layer deposition (ALD) in order to passivate the surface.

\subsection{OMC-Waveguide coupling}

In order to couple light into our optical cavity, we use a coupling waveguide in close proximity to the OMC. The evanescent field of this waveguide overlaps with the optical mode of the cavity and thus, by setting the distance between the two, we can set the external coupling (decay) rate. The cavity decay rate is $\kappa=\kappa\und{i}+\kappa\und{e}$, where $\kappa\und{i}$ is the intrinsic (cold cavity) and $\kappa\und{e}$ is the external decay rate. The efficiency of coupling light into or out of the optical cavity is then given by $\eta\und{dev}=\kappa_e/\kappa$. For a single-sided optical cavity, this efficiency cannot simply be extracted from a reflection spectrum because for a given depth of the cavity dip two solutions exist for $\eta\und{dev}$, where one corresponds to an over-(under-)coupled cavity with $\eta\und{dev,over}>0.5$ ($\eta\und{dev,under}<0.5$). In order to determine whether our device is over- or under-coupled, we employ a technique from~\cite{Groeblacher2013a}, where we stabilize our laser far detuned from the optical cavity and generate sidebands on this carrier using an EOM. We sweep the sideband across the resonance by driving the EOM with a vector network analyzer (VNA) and monitoring the reflected signal at the driving frequency using a high-frequency photodiode. From the resulting amplitude and phase responses, we determine our device to be over-coupled, with $\eta\und{dev} = 0.75$.

This strong over-coupling results in our device being imperfectly sideband resolved $\left(\kappa/4\omega\und{m}\right)^2=0.196<1$. For the present measurement, this does not pose any restrictions, however, as we are not operating in the bad cavity limit either. In particular, when driving at $\omega\und{l}=\omega\und{c}\pm\omega\und{m}$ the optomechanically generated sidebands at $\omega\und{c}\pm 2\omega\und{m}$ are suppressed by 7.8~dB compared to the tone that is resonant with the cavity (at $\omega\und{c}$). In addition, we filter the light leaving the OMC using tuneable Fabry-P\' erot cavities with a bandwidth of 40~MHz, allowing us to further suppress signals that are off-resonant from the optical cavity.

\subsection{Scattering probabilities}

\begin{figure}
\includegraphics[width = .75\columnwidth]{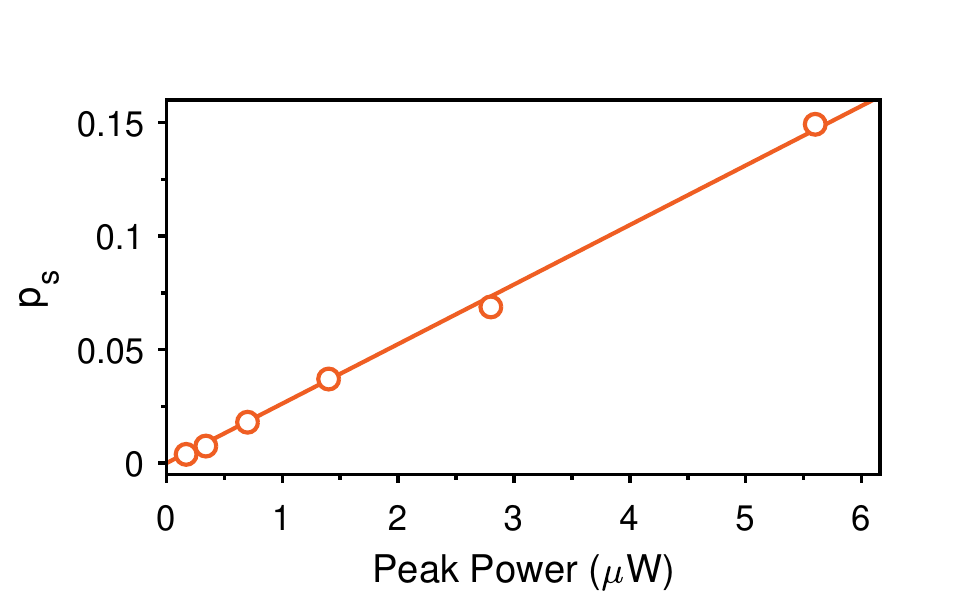}
	\caption{Calibration of scattering probability $p\und{s}$ from input peak power, for 40-ns long pulses. The line is a linear fit to the data. The power that reaches the device is attenuated by the 55\% fiber coupling efficiency. Data set as used for Fig.~\ref{fig2} in the main text.}
	\label{figs1}
\end{figure}

The scattering probability $p\und{s}$ in our experiment describes the probability of either exciting (blue sideband) or reading out (red sideband) the mechanical mode. We retrieve this probability from the measured count rates ($\Gamma\und{r}$ and $\Gamma\und{b}$, respectively) and our calibrated detection efficiency $\eta\und{det}$
\begin{eqnarray}
\Gamma\und{r} = p\und{s,read}\cdot n\und{th}\cdot \eta\und{det},\\
\Gamma\und{b} = p\und{s,write}\cdot (n\und{th}+1)\cdot \eta\und{det},
\end{eqnarray}
where, in the weak coupling limit ($g \ll \kappa$), $p\und{s,r}$ and $p\und{s,b}$ are given by
\begin{eqnarray}
p\und{s,r}=1-\exp\left(\frac{-4\eta\und{dev}g\und{0}^2E\und{p}}{\hbar\omega\und{c}\left(\omega\und{m}^2+\left(\kappa /2\right)^2\right)}\right),\\
p\und{s,b}=\exp\left(\frac{4\eta\und{dev}g\und{0}^2E\und{p}}{\hbar\omega\und{c}\left(\omega\und{m}^2+\left(\kappa /2\right)^2\right)}\right)-1.
\end{eqnarray}
Here, $E\und{p}$ is the total energy of the incident laser pulse~\cite{Hong2017}. In the limit of small scattering probability, where $p\und{s,read/write} \ll 1$, the two values converge to
\begin{equation}
p\und{s} \approx \frac{4\eta\und{dev}g\und{0}^2E\und{p}}{\hbar\omega\und{c}\left(\omega\und{m}^2+\left(\kappa /2\right)^2\right)}.
\end{equation}
Calibration of the detection efficiency $\eta\und{det}$ using an attenuated laser pulse reveals a value of $\eta\und{det} = 0.023$. This value includes the efficiency between the overcoupled optical resonator and the nearby waveguide $\eta\und{dev} = 0.75$ and the coupling efficiency between the waveguide and the fiber $\eta\und{fc} = 0.55$. The remaining losses stem from the filtering setup and the single-photon-detector efficiency. Having calibrated the mode occupation $n\und{th}$, the scattering probability can then be found from the click rate. The calibration between the peak input optical power for 40-ns long pulses and the scattering probability is plotted in Fig.~\ref{figs1}, as a function of the pulse energy, from the same data set as Fig.~\ref{fig2} in the main text. The line is a linear fit to the data, with a proportionality constant of $2.6\times10^{-2}~\mu$W$^{-1}$.

\subsection{Parameters for cross correlation measurement}

In our DLCZ-type pulsed experiment, we send two optical pulses to our device. The first pulse, on the blue sideband of the optomechanical cavity, has a length of 40~ns and a peak power of 25~nW, corresponding to a scattering probability $p\und{s,write}=0.06\%$. The second pulse, detuned to the red sideband, also has a length of 40~ns and a peak power of 750~nW, corresponding to a scattering probability $p\und{s,read}=2\%$. The delay between the two pulses is 150~ns and the repetition rate of the experiment is 25~kHz.

\subsection{Piezoelectric interface}

\begin{figure}
	\includegraphics[width = 1\columnwidth]{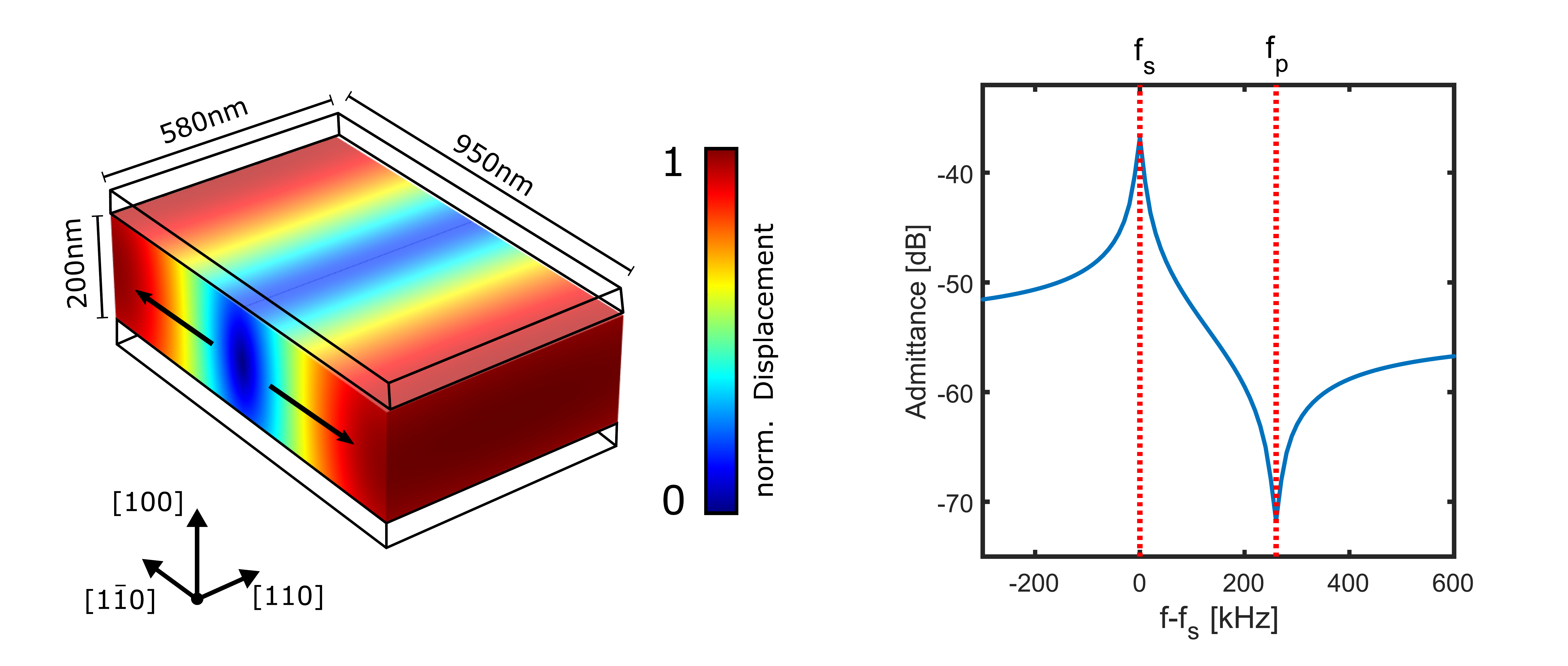}
	\caption{Simulated GaP piezo resonator. The left panel shows a generic block resonator. The thickness is set by the device layer in the main text, while the width is set to closely match the frequency of the confined nanobeam mode and the length chosen to ensure the mode mass matches the nanobeam breathing mode mass. The relation of the block to the crystal axes are displayed. The color coding displays the displacement of the breathing mode along the $[1\bar10]$ direction. Top and bottom electrodes that actuate the mode are indicated by the wire-frame. The right panel shows the simulated electrical admittance between the electrodes. The electromechanical series and parallel resonances, $f\und{s}$ and $f\und{p}$, respectively, are illustrated.}
	\label{figs2}
\end{figure}

Supplementing the nanobeam discussed in the main text with a high-cooperativity piezoelectric interface would allow for transduction between microwave and optical domains. In such a conversion device, the added noise would predominantly be set by the thermal occupation of the nanobeam, which we show to be sufficiently low to allow for conversion of a quantum state. The main figures of merit for a conversion device are the efficiency and the added noise. The latter in particular is critically important to the success of the conversion process, as with sufficiently low noise levels, successful conversion can be heralded through detection events. In particular, the added noise, $N$ follows
 \begin{equation}
 N = \frac{n\und{m}}{\eta\und{e}C\und{em}},
 \end{equation} 
where $\eta\und{e}$ is the external efficiency of the electrical input, $C\und{em}$ the electromechanical cooperativity, and $n\und{m}$ the unwanted population of the mechanical mode, either due to thermal excitation, incoherent photon absorption or unwanted optomechanical excitation. For the largest optomechanical cooperativity measured in Figure~\ref{fig2} in the main text, $C\und{om} = 20$ we find a value oof $n\und{m} = 0.35~\pm~0.01$. If we consider a matched electromechanical cooperativity (peak efficiency would occur at $C\und{em} \approx C\und{om}+1$), and unity external efficiency ($\eta\und{e} = 1$), we would expect to only add $N = 0.02$ photons of noise in the conversion process. A value of $C\und{em}$ that far exceeds the optomechanical cooperativity would further reduce the value of added noise, however at the expense of a reduced conversion efficiency \cite{Wu2019}.

In Fig.~\ref{figs2}, we show a simulated generic GaP piezo-resonator and the corresponding electrical admittance curve. The dimensions of the resonator are chosen to match the mass of the optomechanically coupled mode considered in the main text, such that we maintain a large optomechanical coupling rate. As a best estimate for mechanical dissipation in the system, we use the loss rate of the nanobeam mode ($\gamma\und{m} = 2\pi\times7.96$~kHz, following the mechanical decay rate displayed in Fig.~\ref{fig3}(a) of the main text). The breathing mode of the pictured piezo-resonator is actuated by a vertical electric field, according to the GaP piezoelectric tensor \cite{Nelson1968}, which could be supplied by top-and-bottom electrodes, as is done in reference~\cite{Bochmann2013}. The right panel of Fig.~\ref{figs2} shows the simulated admittance of the resonator, around the value of $f\und{s}$ = 3.05 GHz to approximately match the nanobeam mode. The typical resonance/ anti-resonance curve of the coupled electrical and mechanical system allows us to estimate the piezoelectric coupling coefficient $k\und{eff}^2$, according to $k\und{eff}^2 \approx \left(f\und{p}^2 - f\und{s}^2\right)/f\und{p}^2$, where $f\und{s}$ is the resonant frequency of the mechanical mode, and $f\und{p}$ is the parallel resonance of the coupled electromechanical system. From these values, we extract $k\und{eff}^2 \approx 1.7\times10^{-4}$.

Through the addition of a microwave frequency resonator, we can realize high-cooperativity electromechanical interaction with the piezoelectric resonator. In particular, the cooperativity of the piezoelectric interface is described by $C\und{em} = k\und{eff}^2\omega\und{m}^2/\kappa\und{e}\gamma\und{m}$~\cite{Wu2019}, where $\kappa\und{e}$ is the decay rate of the microwave-frequency resonator, and $\gamma\und{m}$ is the decay rate of the mechanical mode.  Importantly, owing to the small size of the piezoelectric resonator, the addition of a microwave resonator will decrease the value of $k\und{eff}^2$ through additional parasitic capacitance. Taking a relatively large value of 100~fF for the parasitic capacitance (resulting in a characteristic impedance of $520~\Omega$, for comparison, see reference~\cite{Harabula2017}), and the simulated value of 0.19~fF for the capacitance of the piezo-resonator, we estimate a reduced value of $k\und{eff,red}^2 \approx 3.3\times10^{-7}$. Nonetheless, to achieve a cooperativity of $C\und{em} = 20$, the required quality factor for the microwave resonator can be found to be only $Q_{\mu\rm{w}} \equiv \omega\und{m}/\kappa\und{e} \approx 170.$

\end{document}